\documentclass[preprint2]{aastex}

\newcommand {\be} {\begin{equation}}
\newcommand {\ee} {\end{equation}}

\def\spose#1{\hbox to 0pt{#1\hss}}
\def\lta{\mathrel{\spose{\lower 3pt\hbox{$\mathchar"218$}}
        \raise 2.0pt\hbox{$\mathchar"13C$}}}
\def\gta{\mathrel{\spose{\lower 3pt\hbox{$\mathchar"218$}}
        \raise 2.0pt\hbox{$\mathchar"13E$}}}

\shorttitle{The disk vanishes}

\shortauthors{Hameury \& Lasota}

\begin{document}

\title{The disk vanishes}

\author{Jean-Marie Hameury\altaffilmark{1},
and Jean-Pierre Lasota\altaffilmark{2} }

\altaffiltext{1}{Observatoire de Strasbourg, 11 rue de
l'Universit\'e, 67000 Strasbourg, France;
\tt{hameury@astro.u-strasbg.fr}} \altaffiltext{4}{Institut
d'Astrophysique de Paris, UMR 7095 CNRS, Universit\'e Pierre \&
Marie Curie, 98bis Boulevard Arago, 75014 Paris, France;
\tt{lasota@iap.fr}}

\begin{abstract}
Recently published observations of VY Scl stars in their low,
long-intermediate and high states confirm our model \citep{hl02}
according to which accretion disks in such systems must vanish
when their temperature corresponds to thermally unstable
configurations. An observational confirmation of the hypothesis
that the disk disappearance is caused by magnetic effects would
have important consequences for cataclysmic variable evolution
models.
\end{abstract}


\section{INTRODUCTION}

VY Scl stars  are a subclass of Nova-Like (NL) Cataclysmic Variables
(CVs) which show in their long-term light curves the presence of
long low states \citep[see e.g.][]{w95}. The brightness variations
are generally attributed to variations in the rate at which the
secondary star loses matter accreted by the primary white dwarf.
Similar luminosity variations are observed in highly magnetized AM
Her stars -- since these CVs do not have disks these can be only due
to mass-transfer variations. On the other hand, in contrast with AM
Her stars, all VY Scl stars have orbital periods between $\sim 3$
and $\sim 4$ hours.

In their high states VY Scl stars (in which accretion disks are
clearly observed) are among the brightest stationary CVs but their
descent to low states brings them into the range of absolute
magnitudes typical of Dwarf Novae (DN) -- CVs undergoing more or
less regular outbursts -- yet no DN-type outburst from VY Scl stars
has been observed.

There are therefore two VY Scl puzzles. First, the cause of the
observed mass-transfer variations. Second, the reason for the
absence of dwarf-nova type outbursts in the regime of physical
parameters where they are expected. In this article we address only
the second, showing that the solution proposed by \citet[][
hereafter HL02]{hl02} has been strongly confirmed by observations
\citep{ttar,linnell,stan04}. The first puzzle is still unsolved and
is part of the larger puzzle of CV evolution.

\section{No disk, no outburst}

Dwarf nova outbursts are clearly related \citep{smak83} to a thermal
instability occurring when the effective temperature of the
accretion disk corresponds to partial hydrogen ionization, i.e.
$T_{\rm eff} \approx 5800 - 7200$ K \citep[see e.g.][]{l01}. Outside
this instability strip, accretion disks are either in hot or cold
stable equilibria. In such (constant accretion rate) disks the
effective temperature decreases with radius (e.g. the celebrated
``$r^{-3/4}$ law" for hot disks). Therefore, for a given mass
transfer rate, a disk truncated at an inner radius corresponding to
effective temperature $\lta 5800$ K will be cold and stable
\citep{lhh95}. Hence the idea that the absence of outbursts during
the \textsl{low} states of VY Scl stars is due to such truncation.
\citet{leach99} assumed that truncation is due to irradiation of the
inner disk by the hot white dwarf. HL02 showed that such a model is
not viable because it could work only for low-mass and very-hot
white dwarfs, contrary to observations that do not show such a
selection. Instead, HL02 showed that truncation by the magnetic
field of the white dwarf could explain the absence of dwarf nova
outbursts during VY Scl low states. However, HL02 pointed out that
the magnetic moment required to explain the absence of outbursts in
the low state is \textsl{not} sufficient to prevent outbursts during
long \textsl{intermediate} states when VY Scl stars fall to or rise
from the low state. No outbursts are observed during intermediate
states. As explained by HL02, when the characteristic time of
mass-transfer rate variations is longer than the disk's viscous
time, the disk structure follows a sequence of quasi-equilibrium
states and outbursts are unavoidable if the disk temperature
corresponds to the instability strip. The only way to prevent
outbursts is to get rid of the disk: no disk, no outbursts. Or, at
least no dwarf-nova outbursts. The disk must vanish at the latest
when it would enter the instability strip.

HL02 assumed in their calculations that the cause for the disk
disappearance is magnetic: it vanishes when the white dwarf's
magnetospheric radius is approximately equal to the
circularization radius. Numerical simulations give a typical value
of the required magnetic moment as $\gta 7 \times 10^{32} \rm G\
cm^3$.

To summarize: HL02 predict the absence of accretion disks in VY Scl
during the low state and during a substantial fraction of the
intermediate states; when entering the hot stable regime of
mass-transfer rates the disk is gradually re-formed until it gets
back to its full extent in the high state. All these predictions
have been confirmed by observations.

\section{Confrontation with observations}

\citet{ttar} found that the spectra of the VY Scl star TT Ari
observed during a low state show (``virtually") no signs of an
accretion disk. A similar conclusion was arrived to by
\citet{linnell} who found that if the disk were present during the
low state of MV Lyr its effective temperature would have to be less
than 2500 K. This is of course compatible with an absence of a disk
and thus confirms the HL02 predictions. Somewhat paradoxically
\citet{linnell} claim that their conclusion is ``in conflict" with
HL02. This is because they compare their interpretation of IUE
archival observations with the model HL02 use to show that disc
truncation is \emph{not} sufficient to prevent outbursts during long
intermediate states of VY Scl stars. Since MV Lyr is a very slow
riser (few hundred days) this model is \textsl{not} supposed to
apply to these system and there is not contradiction whatsoever
between \citet{linnell} and HL02.

Indeed they also observed MV Lyr in an intermediate state and found
that the spectra can be well represented by a disk extending only
out to half of the tidal truncation radius. A similar conclusion was
reached by \citet{stan04} who observed the eclipsing VY Scl star DW
UMa in an intermediate state. Using eclipse mapping techniques they
found that the luminosity difference between the intermediate and
the high states is almost entirely due to the increase of the disk
radius from $\sim 0.5$ to $\sim 0.75$ of the distance from the white
dwarf to the Roche $L_1$ point. These two observational results are
also in a very good agreement with the HL02 model
\citep{hl05a,hl05b}.

The \citet{linnell} intermediate-state model is isothermal and
truncated at $1.7$ of the white dwarf radius. This radius is much
too small compared to the requirement of HL02 but since according to
the authors the contribution to the flux from innermost annuli is
small this is not very constraining. One should also note that
\citet{linnell} do not take into account the accretion luminosity of
the matter falling onto the white dwarf.

\section{Conclusions}

The properties of VY Scl stars that according to HL02 are required
to avoid dwarf-nova outbursts during low \emph{and} long
intermediate states have been found in at least three binaries of
this type.

HL02 assumed that the disk vanishing is due to the magnetic field of
the white dwarf. The direct evidence of magnetic moments $\gta
10^{32} \rm \ G\ cm^3$ in VY Scl stars is still missing. As
explained in HL02 this evidence is not easy to find. The current
lower limit for detection of magnetic fields in CVs is $\sim 7
\times 10^6$ G \citep{wf00}, which for parameters of MV Lyr, say,
would correspond to $\sim 2.5 \times 10^{33} \rm \  G\ cm^3$.
Confirmation of the magnetic nature of the VY Scl stars would have
important consequences for models of CV evolution since it would
increase the fraction of magnetic systems among the CVs and change
the observed distributions of magnetic vs non-magnetic cataclysmic
binaries \citep[see][]{gan05}. It could support the suggestion of
\citet{regtout} that CV primaries are more magnetic than isolated
white dwarfs because they went through a common envelope phase. In
any case there is growing evidence that binaries containing white
dwarfs with fields $< 7 \times 10^6$~G may constitute the dominant
fraction of the apparently ``non-magnetic" CVs \citep[see
e.g.][]{w04}.

If the quest for magnetic moments in VY Scl stars fails, one would
have to look for other reasons of the disk vanishing. Accretion
disks cannot be present during most of the intermediate state
duration. If they were they would go into outburst. Not because of
the prediction of the disk instability model \citep[see][for a
review]{l01}, but because CVs with very similar binary parameter do
so when they have the corresponding absolute magnitudes.

\acknowledgments
We thank Al Linnell for an interesting e-mail
communication.

\end{document}